\documentclass[aps,pre,reprint,amsmath,amssymb,longbibliography,lengthcheck,showpacs,superscriptaddress,floatfix]{revtex4-2}
\usepackage{graphicx}
\usepackage{dcolumn}
\usepackage{bm}
\usepackage{color}
\def\bra#1{\langle #1|}
\def\ket#1{|#1\rangle}

\begin{document}
%
\title{Unified study of viscoelasticity and sound damping in hard and soft amorphous solids}
%
\author{Hideyuki Mizuno}
\email{hideyuki.mizuno@phys.c.u-tokyo.ac.jp; contributed equally}
\affiliation{Graduate School of Arts and Sciences, The University of Tokyo, Tokyo 153-8902, Japan}
\author{Kuniyasu Saitoh}
\email{k.saitoh@cc.kyoto-su.ac.jp; contributed equally}
\affiliation{Department of Physics, Faculty of Science, Kyoto Sangyo University, Kyoto 603-8555, Japan}
\author{Yusuke Hara}
\affiliation{Graduate School of Arts and Sciences, The University of Tokyo, Tokyo 153-8902, Japan}
\author{Atsushi Ikeda}
\affiliation{Graduate School of Arts and Sciences, The University of Tokyo, Tokyo 153-8902, Japan}
\affiliation{Research Center for Complex Systems Biology, Universal Biology Institute, The University of Tokyo, Tokyo 153-8902, Japan}
%
\date{\today}
%
\begin{abstract}
Recent research has made significant progress in understanding the non-phonon vibrational states present in amorphous materials.
It has been established that their vibrational density of states follows non-Debye scaling laws.
Here, we show that the non-Debye scaling laws play a crucial role in determining material properties of a broad range of amorphous solids, from ``hard" amorphous solids like structural glasses to ``soft" amorphous solids such as foams and emulsions.
We propose a unified framework of viscoelasticity and sound damping for these materials.
Although these properties differ significantly between hard and soft amorphous solids, they are determined by the non-Debye scaling laws.
We also validate our framework using numerical simulations.
\end{abstract}
%
\maketitle
%
\section{Introduction}
%
Amorphous materials exist in various forms, such as structural glasses, granular materials, foams, emulsions, and biological systems~\cite{Rheology,Bonn_2017,Janssen_2019}, which are made up of densely packed constituents in disordered states.
Recent studies have established that amorphous materials universally exhibit non-phonon vibrational states, which are observed as excess vibrational states over the Debye prediction, referred to as boson peak~(BP)~\cite{Buchenau_1984,Yamamuro_1996,Beltukov_2016,Mori_2020}, and as quasi-localized vibrational states in the low-frequency, continuum limit~\cite{Mizuno_2017,Shimada_2018,Wang_2019}.
The vibrational density of states~(vDOS) $g(\omega)$~(where $\omega$ is frequency) follows the non-Debye scaling laws: $g(\omega) \propto \omega^2$ in the BP regime~\cite{Charbonneau2016,Shimada_2020} and $g(\omega) \propto \omega^{d+1}$~(where $d$ is the spatial dimension) in the continuum limit regime~\cite{Mizuno_2017,Shimada_2018,Wang_2019}.
The non-phonon vibrational states play a crucial role in determining material properties such as mechanical and thermal properties~\cite{Wyart_2005,schirmacher_2006,Schirmacher_2007,Wyart_2010,Degiuli_2014}.

Here, we bring our attenuation to viscoelasticity and sound wave propagation.
In structural glasses, the loss modulus $G''$ and the sound attenuation $\Gamma$ stem from the amorphous structures.
Many previous simulations~\cite{Monaco_2009,Marruzzo_2013,Mizuno_2014,Mizuno_2018,Moriel_2019,Wang2_2019} and experiments~\cite{ruffle_2006,Masciovecchio_2006,Monaco2_2009,Baldi_2010,Baldi_2011,Baldi_2016} have established that sound waves exhibit viscous damping $\Gamma \propto \omega^2$ in the BP regime and Rayleigh scattering $\Gamma \propto \omega^{d+1}$ in the continuum limit regime.
These behaviors are intimately linked to the non-Debye scaling laws through the relation $\Gamma \sim g(\omega)$, which is described by the generalized Debyel model~\cite{Marruzzo_2013,Mizuno_2018}.

On the other hand, previous theoretical~\cite{Liu_1996,Tighe_2011,Baumgarten_2017,Hara_2023,Hara2_2023} and experimental~\cite{Mason_1995,Cohen-Addad_1998,Hebraud_2000,Gopal_2003,Besson_2008,Marze_2008,Kropka_2010,Krishan_2010,Conley_2019,Hara2_2023} works have investigated different types of amorphous materials, such as foams and emulsions, which are considered strongly damped solids due to the presence of viscous forces.
In particular, Refs.~\cite{Tighe_2011,Baumgarten_2017} have researched macrorheology and formulated the complex modulus based on vibrational eigenvectors and the vDOS.
Similarly, Ref.~\cite{Hara_2023} has explored microrheology and developed the corresponding complex modulus.
The non-Debye scaling laws control both macrorheology and microrheology, as they govern the sound damping in structural glasses.
Moreover, many experimental studies~\cite{Mason_1995,Cohen-Addad_1998,Hebraud_2000,Gopal_2003,Besson_2008,Marze_2008,Kropka_2010,Krishan_2010,Conley_2019} have observed the loss modulus following $G'' \propto \omega^{1/2}$, which is called the anomalous viscous loss~\cite{Liu_1996}.
A most recent work~\cite{Hara2_2023} demonstrated a direct connection between the anomalous viscous loss and the non-Debye scaling law $g(\omega) \propto \omega^2$~(boson peak) through theoretical and experimental approaches.

In this work, we present a unified framework of viscoelasticity and sound wave propagation for amorphous materials, including ``hard" solids like structural glasses and ``soft" solids like foams and emulsions.
We consider the simplest model of amorphous solids, which consists of randomly jammed particles of mass $m$, interacting through the harmonic potential~\cite{Hern_2003,Hecke_2010};
\begin{equation}~\label{pot-simple}
\phi(r) = \frac{k}{2} \left( \sigma - r \right)^2 H(\sigma -r),
\end{equation}
where $\sigma$ is the diameter of the particles, $k$ is the stiffness, and $H(r)$ is the Heaviside step function.
The static packings of the system undergo the jamming transition at the transition density~$\phi_J$, becoming isostatic with the contact number per particle $z$ being equal to $2d$.
Near the transition, physical quantities such as the excess contact number $\Delta z = z-2d$ and the elastic moduli follow power-law scalings with the excess density $\Delta \phi = \phi - \phi_J$, or equivalently, the pressure $p \propto \Delta \phi$~\cite{Hern_2003,Hecke_2010}.
The non-Debye scaling laws of the vDOS $g(\omega)$ also exhibit the critical behavior~(as in Eq.~(\ref{eq.dos}))~\cite{Charbonneau2016,Shimada_2020,Mizuno_2017}.
We aim to establish scaling laws of complex modulus and sound attenuation for both hard and soft amorphous solids by utilizing the non-Debye scaling laws.

\section{Formulation}
\textit{Basic formalism---.}
Our formulation is based on the previous works~\cite{Tighe_2011,Baumgarten_2017}, where the complex shear modulus $G_\ast(\omega)$ is formulated as follows.
To begin with, we introduce the ``extended" displacement vector, which includes the amplitude of the shear strain $\epsilon$, as $\ket{\vec{u}_\text{ex}(t)} = \left[ \vec{u}_1(t),\cdots,\vec{u}_N(t), \sigma \epsilon(t) \right]$~($dN+1$-dimensional vector) and its Fourier transformation $\ket{\tilde{u}_\text{ex}(\omega)}$, where $N$ denotes the number of particles.
We then deal with Newton's equation of motion with a force acting along the $\sigma \epsilon$-coordinate;
\begin{equation} \label{eqmotionfr}
\left( \mathcal{M}_\text{ex} - m\omega^2 + i\omega \mathcal{C}_\text{ex}\right) \ket{\tilde{u}_\text{ex}(\omega)} =  \frac{L^d}{\sigma} \tilde{\tau}_\text{sh}(\omega) \ket{\gamma},
\end{equation}
where $\mathcal{M}_\text{ex}$ and $\mathcal{C}_\text{ex}$ are respectively the ``extended" Hessian and damping matrices.
$\tilde{\tau}_\text{sh}(\omega)$ is the imposed shear stress, $L$ is the system length, and $\ket{\gamma}=[ \vec{0},\cdots,\vec{0}, 1]$.
Please refer to Appendix~\ref{app.hessian} for detailed descriptions of Eq.~(\ref{eqmotionfr}) with $\mathcal{M}_\text{ex}$ and $\mathcal{C}_\text{ex}$.

The Hessian matrix is expanded with its eigenvalues $m\omega_{\text{ex} k}^2$ and eigenvectors $\ket{\vec{e}_{\text{ex} k}}$ as $\mathcal{M}_\text{ex} = \sum_{k=1}^{dN+1} m\omega_{\text{ex} k}^2 \ket{\vec{e}_{\text{ex} k}} \bra{\vec{e}_{\text{ex} k}}$.
For the damping matrix, we suppose Stokes dissipation as $\mathcal{C}_\text{ex} = \zeta I_{dN+1} = \sum_{k=1}^{dN+1} \zeta \ket{\vec{e}_{\text{ex} k}} \bra{\vec{e}_{\text{ex} k}}$, where $\zeta$ denotes strength of the dissipation, and $I_{dN+1}$ is $(dN+1)\times (dN+1)$ unit matrix.
We then get
\begin{equation} \label{eqgastoriginal}
\frac{1}{G_\ast (\omega)} = \frac{\sigma^{-1} \bra{\gamma} \tilde{u}_\text{ex}(\omega) \rangle }{\tilde{\tau}_\text{sh}(\omega)}
= \sum_{k=1}^{dN+1} \frac{L^d \sigma^{-2} |\bra{\vec{e}_{\text{ex} k}} \gamma \rangle|^2 }{m(\omega_{\text{ex} k}^2 -\omega^2) + i\omega \zeta}.
\end{equation}
Following Refs.~\cite{Tighe_2011,Baumgarten_2017}, we assume that $|\bra{\vec{e}_{\text{ex} k}} \gamma \rangle|$ does \textit{not} depend on the eigenmode $k$, and obtain
\begin{equation} \label{eqgast}
\frac{1}{G_\ast (\omega)} \sim \int_0^\infty d\omega' \frac{1 }{m(\omega'^2 -\omega^2) + i\omega \zeta} g(\omega').
\end{equation}
In Eq.~(\ref{eqgast}), we should put the vDOS $g_\text{ex}(\omega)$, \textit{i.e.,} distribution of $\omega_{\text{ex} k}$.
However, one additional degree of freedom $\sigma \epsilon$ makes a tiny variation on the vDOS, and we deal with $g(\omega)$, \textit{i.e.,} distribution of $\omega_k$ from the ``usual" Hessian matrix $\mathcal{M}$~(without the shear degree of freedom).

We note that Stokes dissipation is applicable not only to structural glasses but also to foams and emulsions.
For instance, in a two-dimensional~($d=2$) model of foams proposed by Durian~\cite{Durian_1995,Durian_1997}, contacting bubbles experience a viscous force $\vec{f}_{\text{visc}} = -\mu \Delta \vec{v}$ which opposes their relative velocity $\Delta \vec{v}$, where $\mu$ is the viscosity.
In this model, $\mathcal{C}_\text{ex}$ can be approximated as Stokes dissipation, where $\mathcal{C}_\text{ex} = \zeta I_{2N+1}$ and $\zeta = \mu z$ with $z \approx 2d = 4$~(see Appendix~\ref{app.damping}).

Using $G_\ast (\omega)$ in Eq.~(\ref{eqgast}), the sound speed $c_T(\omega)$ and the attenuation rate $\Gamma_T(\omega)$ of the transverse shear wave are determined through the continuum mechanics as~\cite{continuum}
\begin{equation} \label{eq.sound}
\begin{aligned}
c_T(\omega)^{-1} &= \text{Re} \left[ \sqrt{\frac{\rho}{G_\ast(\omega)}} \right], \\
-\frac{\Gamma_T(\omega)}{\omega c_T(\omega)} &= \text{Im} \left[ \sqrt{\frac{\rho}{G_\ast(\omega)}} \right].
\end{aligned}
\end{equation}
Eqs.~(\ref{eqgast}) and~(\ref{eq.sound}) are basic formulations for viscoelasticity and sound wave propagation, respectively.
There are two key parameters; the vDOS $g(\omega)$ and the dissipation $\zeta$.

\textit{vDOS $g(\omega)$---.}
The vDOS of the present system~(\ref{pot-simple}) has been understood, which is given as~\cite{Charbonneau2016,Shimada_2020,Mizuno_2017};
\begin{equation} \label{eq.dos}
g(\omega)
\sim
\left\{ 
\begin{aligned}
& 1 & \left( \omega_\ast < \omega < \sqrt{k/m} \right), \\
& \left( \frac{\omega}{\omega_\ast} \right)^2 & (\omega_0 < \omega < \omega_\ast), \\
& \left( \frac{\omega}{\omega_\ast} \right)^{d+1} & (\omega < \omega_0),
\end{aligned} 
\right.
\end{equation}
where $\omega_\ast$ and $\omega_0$~($\omega_0 \ll \omega_\ast$) are characteristic frequencies both of which follow the same scaling law as $\omega_\ast \propto \omega_0 \propto \Delta \phi^{1/2} \propto p^{1/2}$~\cite{Mizuno_2017}.
$g(\omega) \sim (\omega/\omega_\ast)^2$ at $\omega_0 < \omega < \omega_\ast$ is the non-Debye scaling law~(independent of $d$).
In much lower frequency~(continuum limit) regime of $\omega < \omega_0~(\ll \omega_\ast)$, there exists another non-Debye scaling law $g(\omega) \sim (\omega/\omega_\ast)^{d+1}$ in $d=2$ and $3$ dimensions~\cite{Mizuno_2017,Shiraishi_2023}.

\textit{Dissipation $\zeta$---.}
The dissipation strength is an important factor that distinguishes hard and soft amorphous solids.
In structural glasses at low temperatures~\cite{Monaco_2009,Marruzzo_2013,Mizuno_2014}, the dissipation is usually small.
Therefore, we apply the small dissipation condition;
\begin{equation} \label{smallviscosity}
\frac{\zeta}{m \omega} \ll 1.
\end{equation}

On the other hand, particles are highly damped in foams and emulsions due to the viscous forces~\cite{Durian_1995,Durian_1997}.
Hence, we implement the large dissipation condition;
\begin{equation} \label{largeviscosity}
\frac{\zeta}{m \omega} \gg 1.
\end{equation}
It is worth emphasizing that $g(\omega)$ is a common factor between hard and soft amorphous solids, whereas $\zeta$ differs between them as in Eqs.~(\ref{smallviscosity}) and~(\ref{largeviscosity}).

\section{Theoretical predictions}
\textit{Hard amorphous solids---.}
Calculating $G_\ast(\omega)$ in Eq.~(\ref{eqgast}) with applying Eq.~(\ref{eq.dos}) for $g(\omega)$ and Eq.~(\ref{smallviscosity}) for $\zeta$, we obtain the scaling laws of $G_\ast$;
\begin{equation} \label{hardvisco}
\frac{G_\ast}{\omega_\ast}
\sim
\left\{ 
\begin{aligned}
& 1 + i \left(\frac{\omega}{\omega_\ast} \right) & \left( \frac{\omega}{\omega_\ast} \gg 1 \right), \\
& 1 + i \left(\frac{\omega}{\omega_\ast} \right) & \left( \frac{\omega_0}{\omega_\ast} \ll \frac{\omega}{\omega_\ast} \ll 1 \right), \\
& 1 + i \left(\frac{\omega}{\omega_\ast} \right)^{d} & \left( \frac{\omega}{\omega_\ast} \ll \frac{\omega_0}{\omega_\ast} \right).
\end{aligned} 
\right.
\end{equation}
Note that taking the limit of $\omega \to 0$ in Eq.~(\ref{hardvisco}) produces $G_\ast \sim \omega_\ast \propto \Delta \phi^{1/2} \propto p^{1/2}$ for the static shear modulus~\cite{Hern_2003,Hecke_2010}.
We then get $c_T$ and $\Gamma_T$ in Eq.~(\ref{eq.sound});
\begin{equation} \label{hardsound}
\begin{aligned}
\frac{c_T}{\omega_\ast^{1/2}}
&\sim
\left\{
\begin{aligned}
& \left(\frac{\omega}{\omega_\ast} \right)^{1/2} & \left( \frac{\omega}{\omega_\ast} \gg 1 \right),\\
& 1 & \left( \frac{\omega}{\omega_\ast} \ll 1 \right),
\end{aligned}
\right.
\\
\frac{\Gamma_T}{\omega_\ast}
&\sim
\left\{
\begin{aligned}
& \left(\frac{\omega}{\omega_\ast} \right) & \left( \frac{\omega}{\omega_\ast} \gg 1 \right),\\
& \left(\frac{\omega}{\omega_\ast} \right)^2 & \left( \frac{\omega_0}{\omega_\ast} \ll \frac{\omega}{\omega_\ast} \ll 1 \right),\\
& \left(\frac{\omega}{\omega_\ast} \right)^{d+1} & \left( \frac{\omega}{\omega_\ast} \ll \frac{\omega_0}{\omega_\ast} \right).
\end{aligned} 
\right.
\end{aligned}
\end{equation}
Eq.~(\ref{hardsound}) matches numerical observations on structural glasses in a precise manner as reported in Ref.~\cite{Mizuno_2018}.
At high frequencies $\omega/\omega_\ast \gg 1$, our predictions are in agreement with the effective medium theory~\cite{Wyart_2010, Degiuli_2014}.
Therefore, our formulations are able to accurately explain the behavior of structural glasses.
Specifically, the imaginary part of $G_\ast$, $G''$, and $\Gamma_T$~(at $\omega/\omega_\ast \ll 1$) are closely related to the non-Debye scaling laws, which is consistent with the generalized Debyel model as described in Refs.~\cite{Marruzzo_2013,Mizuno_2018}.

\textit{Soft amorphous solids---.}
Applying Eq.~(\ref{largeviscosity}) for $\zeta$ instead of Eq.~(\ref{smallviscosity}), we obtain $G_\ast$ as
\begin{equation} \label{softvisco}
\frac{G_\ast}{\omega_\ast}
\sim
\left\{
\begin{aligned}
& \left( \frac{\omega\zeta}{m\omega_\ast^2} \right)^{1/2} + i \left( \frac{\omega\zeta}{m\omega_\ast^2} \right)^{1/2} & \left( \frac{\omega\zeta}{m\omega_\ast^2} \gg 1 \right), \\
& 1 + i \left( \frac{\omega\zeta}{m\omega_\ast^2} \right)^{1/2} & \left( \frac{\omega_0^2}{\omega_\ast^2} \ll \frac{\omega\zeta}{m\omega_\ast^2} \ll 1 \right), \\
& 1 + i \left( \frac{\omega\zeta}{m\omega_\ast^2} \right) & \left(\frac{\omega\zeta}{m\omega_\ast^2} \ll \frac{\omega_0^2}{\omega_\ast^2} \right).
\end{aligned}
\right.
\end{equation}
We then get $c_T$ and $\Gamma_T$ as
\begin{equation} \label{softsound}
\begin{aligned}
\frac{c_T}{\omega_\ast^{1/2}}
&\sim
\left\{ 
\begin{aligned}
&  \left(\frac{\omega\zeta}{m\omega_\ast^{2}} \right)^{1/4} &\left (\frac{\omega\zeta}{m\omega_\ast^{2}} \gg 1 \right), \\
&  1  & \left( \frac{\omega\zeta}{m\omega_\ast^{2}} \ll 1 \right), 
\end{aligned} 
\right. \\
\frac{\Gamma_T \zeta}{m\omega_\ast^{2}}
&\sim
\left\{ 
\begin{aligned}
&  \frac{\omega\zeta}{m\omega_\ast^{2}} & \left(\frac{\omega\zeta}{m\omega_\ast^{2}} \gg 1 \right), \\
&  \left(\frac{\omega\zeta}{m\omega_\ast^{2}} \right)^{3/2} & \left( \frac{\omega_0^2}{\omega_\ast^2} \ll \frac{\omega\zeta}{m\omega_\ast^{2}} \ll 1 \right), \\
&  \left(\frac{\omega\zeta}{m\omega_\ast^{2}} \right)^{2} & \left( \frac{\omega\zeta}{m\omega_\ast^{2}} \ll \frac{\omega_0^2}{\omega_\ast^2} \right).
\end{aligned} 
\right.
\end{aligned}
\end{equation}
The scaling laws presented in Eq.~(\ref{softvisco}) are consistent with those of the microrheology established in Refs.~\cite{Hara_2023,Hara2_2023}.
The behavior of $G'' \propto \omega^{1/2}$ at ${\omega\zeta}/(m\omega_\ast^2) \gg {\omega_0^2}/{\omega_\ast^2}$ is the anomalous viscous loss~\cite{Liu_1996} which has been observed by many experiments~\cite{Mason_1995,Cohen-Addad_1998,Hebraud_2000,Gopal_2003,Besson_2008,Marze_2008,Kropka_2010,Krishan_2010,Conley_2019,Hara2_2023}.
Note that the dynamics of soft amorphous solids are often modeled using overdamped dynamics~\cite{Tighe_2011, Baumgarten_2017, Hara_2023, Hara2_2023, Durian_1995, Durian_1997}.
The scaling laws of $G_\ast$ in Eq.~(\ref{softvisco}) remain unchanged even if we consider overdamped dynamics~(see Appendix~\ref{app.over}).

Significant differences exist in the scaling behaviors of the complex modulus and sound damping between hard and soft amorphous solids, as evident in Eqs.~(\ref{hardvisco}) to~(\ref{softsound}).
Structural glasses, being hard solids, exhibit $\Gamma_T \propto \omega^2$~(viscous damping) and $\Gamma_T \propto \omega^{d+1}$~(Rayleigh scattering), which are closely related to the non-Debye scaling laws as $\Gamma_T \sim g(\omega)$~\cite{Marruzzo_2013,Mizuno_2018}.
On the other hand, foams and emulsions, being soft solids, exhibit a totally different scattering law of $\Gamma_T \propto \omega^{3/2}$ with a nontrivial exponent of $3/2$.
This behavior arises from a coupling between the non-Debye scaling laws and strong dissipation.
In the remainder of the paper, we will demonstrate Eq.~(\ref{softsound}) of soft solids by conducting molecular dynamics~(MD) simulations on a foam model.

\section{Numerical simulation on foams}
%
We consider a foam model~\cite{Durian_1995,Durian_1997} made up of $N$ disk particles with mass $m$ inside a $L\times L$ square periodic box in two-dimensional space ($d=2$).
The particles interact through elastic and repulsive forces, which are defined in Eq.~(\ref{pot-simple}), \textit{i.e.}, $f_{\text{el}}=k \delta$ where $\delta = \sigma - r~(>0)$ represents the overlap between the particles.
In addition, the particles that come into contact experience a viscous force $\vec{f}_{\text{visc}}=-\mu \Delta \vec{v}$ with the viscosity $\mu$, which opposes their relative velocity $\Delta \vec{v}$. 

We carried out MD simulations following Refs.~\cite{Saitoh_2019,Saitoh_2021,MizunoIkeda2022}.
Our system is a $50\%:50\%$ binary mixture of $N=2097152$ particles, where the two species have the same mass $m$ but different diameters $\sigma_S$ and $\sigma_L=1.4\sigma_S$.
The packing fraction, $\phi=N \pi (\sigma_L^2+\sigma_S^2)/8L^2$, is set to be greater than the jamming density $\phi_J\simeq 0.8433$~\cite{Hern_2003,Hecke_2010}.
To generate the reference configuration of the system, denoted by $[\vec{r}_1,\dots,\vec{r}_N]$, we used the FIRE algorithm~\cite{Bitzek_2006} to minimize the total potential energy of the system $E=\sum_{i<j} \phi(r_{ij})$.
In the following, we measure mass, length, and time in terms of $m$, $\sigma_0 = (\sigma_L+\sigma_S)/2$, and $t_0 = \sqrt{m/k}$, respectively.

To excite the transverse shear wave at the initial time, we set the velocity vector $\vec{v}_i$ as
\begin{equation}~\label{eq:initial}
\vec{v}_i(t=0) = \frac{d\vec{u}_i}{dt}(t=0) =\vec{P}_T \sin(\vec{q}\cdot\vec{r}_i),
\end{equation}
where $\vec{q}$ is the wave vector, and $\vec{P}_T$ is the polarisation vector that is perpendicular to $\vec{q}$ as $\vec{P}_T \cdot\vec{q}=0$.
Starting with $\vec{u}_i(t=0)=\vec{0}$, we solve the equation of motion at $t>0$;
\begin{equation}~\label{eq:motion}
m\frac{d^2\ket{\vec{u}(t)}}{dt^2} = -\mathcal{M} \ket{\vec{u}(t)} - \mathcal{C} \frac{d\ket{\vec{u}(t)}}{dt},
\end{equation}
where $\ket{\vec{u}(t)} = \left[ \vec{u}_1(t),\cdots,\vec{u}_N(t) \right]$~(without the shear degree of freedom), and $\mathcal{M}$ and $\mathcal{C}$ are the ``usual" Hessian and damping matrices~(see also Appendix~\ref{app.hessian}).
Note that our simulations are in the harmonic approximation limit, thus excluding any anharmonic effects due to \textit{e.g.}, opening and closing contacts \cite{Schreck_2011,Deen_2014,Saitoh_2015,Deen_2016}.
The viscosity $\mu$ in $\mathcal{C}$ introduces a microscopic time scale $t_d = \mu/k$.
We fix the ratio, $t_d/t_0 = \mu/\sqrt{mk}$, to be unity~\cite{Saitoh_2021}, so that the large dissipation condition~(\ref{largeviscosity}) is satisfied.
The control parameter is then $\phi$~($>\phi_J$), or equivalently, $p \propto \Delta \phi$.

We analyze the time correlation function,
\begin{equation}
C_T(q,t) = \frac{\langle \vec{v}_T(\vec{q},t) \cdot \vec{v}_T(-\vec{q},0) \rangle}{\langle |\vec{v}_T(\vec{q},0)|^2\rangle},
\label{eq:VAF}
\end{equation}
of the Fourier-transformed, transverse velocity,
\begin{equation}
\vec{v}_T(\vec{q},t)=\sum_{j=1}^N \left\{ \vec{v}_j(t)- \left[ \vec{q} \cdot \vec{v}_j(t) \right] \frac{\vec{q}}{q} \right\} e^{-i\vec{q}\cdot\vec{r}_j},
\end{equation}
where $q = |\vec{q}|$ is the wavenumber, and $\langle \rangle$ denotes ensemble average.
As the initial standing wave~(described in Eq.~(\ref{eq:initial})) is attenuated, $C_T(q,t)$ decays with time, the behavior of which is fitted to the damped harmonic oscillator model~\cite{Mizuno_2018,Saitoh_2019,Saitoh_2021}~(see Fig.~\ref{fig:correlation} in Appendix~\ref{app.timecor});
\begin{equation} \label{eq:dhom}
C_T(q,t) = e^{-\Gamma_T(q)t}\cos\left[ \Omega_T(q)t \right].
\end{equation}
This fitting procedure provides data for $\Omega_T(q)$, the frequency, and $\Gamma_T(q)$, the attenuation rate.
The sound speed is then calculated by $c_T(q) = \Omega_T(q)/q$.

\begin{figure}[t]
\centering
\includegraphics[width=\columnwidth]{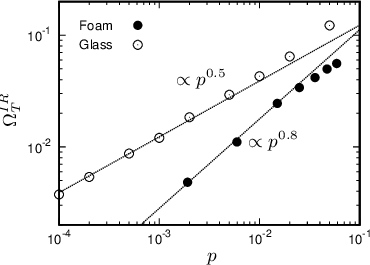}
\caption{ \label{fig:irr}
The Ioffe-Regel limit in the foams.
We plot $\Omega_T^{IR}$ against the pressure $p$.
We also present data on structural glasses reported in Ref.~\cite{Mizuno_2018}.
The dotted lines show the scaling laws of $\Omega_T^{IR} \propto p^{0.8}$ for the foam and $\Omega_T^{IR} \propto p^{0.5}$ for the glass.
}
\end{figure}

\textit{Ioffe-Regel limit---.}
We begin by calculating the Ioffe-Regel frequency, denoted as $\Omega_T^{IR}$, which is given by the formula ${\pi\Gamma_T}/{\Omega_T^{IR}} = 1$. This frequency sets a limit on the maximum propagation frequency~\cite{Mizuno_2018,Saitoh_2019,Saitoh_2021}.
Our findings on $\Omega_T^{IR}$ are presented in Fig.~\ref{fig:irr}.
In this figure, we also show data on the glass model extracted from Ref.~\cite{Mizuno_2018}.
Our results indicate that for a fixed pressure, $\Omega_T^{IR}$ is much lower in the foam model than in the glass model.
This can be explained by the fact that in foams, attenuation arises not only from amorphous structures but also from viscous forces.
In both foams and glasses, $\Omega_T^{IR}$ approaches zero frequency as the jamming transition is approached.
For glasses~\cite{Mizuno_2018}, we observe the scaling law $\Omega_T^{IR} \propto p^{0.5}$, whereas for foams, $\Omega_T^{IR}$ decreases more sharply and follows $\Omega_T^{IR} \propto p^{0.8}$.

\begin{figure}[t]
\includegraphics[width=\columnwidth]{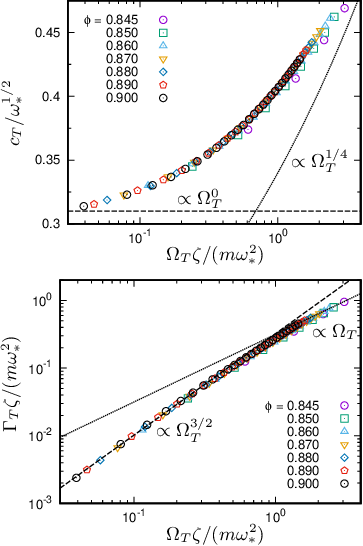}
\caption{ \label{fig:originalscale}
Transverse shear waves in the foams.
We plot (a) $c_T/\omega_\ast^{1/2}$ and (b) $\Gamma_T \zeta/(m\omega_\ast^2)$ as a function of $\Omega_T \zeta/(m\omega_\ast^2)$ for several different $\phi$.
The dashed lines indicate $c_T \propto \Omega_T^{0}$ and $\Gamma_T \propto \Omega_T^{3/2}$ for $\Omega_T \zeta/(m\omega_\ast^2) \ll 1$, while the dotted lines indicate $c_T \propto \Omega_T^{1/4}$ and $\Gamma_T \propto \Omega_T$ for $\Omega_T \zeta/(m\omega_\ast^2) \gg 1$.
}
\end{figure}

\textit{Sound speed and attenuation rate---.}
We now present the values of $c_T$ and $\Gamma_T$ as functions of $\Omega_T$.
In Appendix~\ref{app.sound}, we provide raw data in Fig.~\ref{fig:original}.
Then, to verify our theoretical prediction in Eq.~(\ref{softsound}), Fig.~\ref{fig:originalscale} presents scaled data on $c_T/\omega_\ast^{1/2}$ and $\Gamma_T \zeta/(m\omega_\ast^2)$ as functions of $\Omega_T \zeta/(m\omega_\ast^2)$.
Here, we use $\zeta = \mu z \approx 4$, and the values of $\omega_\ast = 1.58 p^{1/2}$ are taken from Ref.~\cite{Mizuno_2017}.
The scale data collapse onto a single curve regardless of the density values $\phi$, which verifies the theoretical prediction in Eq.~(\ref{softsound}).
In particular, we recognize that $\Gamma_T$ varies as $\Omega_T^{3/2}$ at lower $\Omega_T \zeta/(m\omega_\ast^2) \ll 1$, with a nontrivial exponent of $3/2$.
This scaling law crosses over to $\Gamma_T \propto \Omega_T$ at higher $\Omega_T \zeta/(m\omega_\ast^2) \gg 1$. 
Eq.~(\ref{softsound}) predicts that $\Gamma_T \propto \Omega_T^2$ at much lower $\Omega_T \zeta/(m\omega_\ast^2) \ll \omega_0^2/\omega_\ast^2 \simeq 4 \times 10^{-3}$.
However, the present data do not access this regime.

\section{Conclusion}
%
In summary, we have presented a unified framework explaining the viscoelasticity and sound wave propagation for hard and soft amorphous solids.
Our framework, which is outlined in Eqs.~(\ref{eqgast}) and~(\ref{eq.sound}), is based on two key parameters; the vDOS $g(\omega)$ and the dissipation $\zeta$.
The vDOS is common to both types of amorphous solids and is determined by their amorphous structures.
However, the dissipation is what distinguishes between hard and soft solids.
Hard solids are characterized by the small dissipation condition (Eq.~(\ref{smallviscosity})), while soft solids have the large dissipation condition (Eq.~(\ref{largeviscosity})).
Our formulations predict the scalings of the complex shear modulus and sound wave propagation, as given by Eqs.~(\ref{hardvisco}) through~(\ref{softsound}).
These predictions are in agreement with numerical simulation results for structural glasses reported in Ref.~\cite{Mizuno_2018} and foams studied in this work.

In the case of small dissipation~(structural glasses), we observe viscous damping, $\Gamma_T \propto \omega^2$, and Rayleigh scattering, $\Gamma_T \propto \omega^{d+1}$.
These phenomena come from non-Debye scaling laws of $g(\omega)$.
However, in the case of large dissipation~(foams), we observe a completely different scattering behavior, $\Gamma_T \propto \omega^{3/2}$, with a nontrivial exponent of $3/2$.
This behavior is attributed to a coupling of two distinct effects from non-Debye scaling laws and dissipation.
Interestingly, this coupling can also occur in structural glasses at \textit{finite} temperatures.
We have observed that Rayleigh scattering, $\Gamma_T \propto \omega^{d+1}$, is dominated by $\Gamma_T \propto \omega^{3/2}$ due to anharmonic effects, as reported by previous simulations~\cite{Mizuno_2019,Mizuno_2020,Wang_2020}.
This phenomenon can be attributed to the shift from the small dissipation condition to the large dissipation condition~\cite{Marruzzo2_2013}, but further investigation is needed to verify this point.
In the future, it is interesting to describe the present results in terms of the other theoretical frameworks, including the effective medium theory or fluctuating elasticity theory~\cite{schirmacher_2006,Schirmacher_2007,Wyart_2010,Degiuli_2014,Caroli_2019,Shimada2_2020,Kapteijns_2021,Mahajan_2021}, the random matrix approach~\cite{Conyuh_2021,Vogel_2023,Philipp_2024}, and other microscopic approaches~\cite{Damart_2017,Baggioli_2022,Szamel_2022}.

For structural glasses, acoustic properties of sound speed and attenuation rate have been measured experimentally with light, inelastic X-ray, and neutron scattering techniques~\cite{ruffle_2006,Masciovecchio_2006,Monaco2_2009,Baldi_2010,Baldi_2011,Baldi_2016}.
Meanwhile, for foams and emulsions, various macrorheology and microrheology experimental techniques have been used to measure the mechanical properties of complex elastic moduli~\cite{Hara2_2023,Mason_1995,Cohen-Addad_1998,Hebraud_2000,Gopal_2003,Besson_2008,Marze_2008,Kropka_2010,Krishan_2010,Conley_2019}.
Our framework can unify these different experimental measurements for the two different solids based on the vDOS and dissipation.

\subsection*{Acknowledgments}
We thank B. P. Tighe, K. Taghizadeh, V. Magnanimo, H. Cheng, and S. Luding for fruitful discussions.
This work was supported by JSPS KAKENHI Grant Numbers 18H05225, 19H01812, 20H00128, 20H01868, 21H01006, 22K03459, 22K03543, 23H04495.
This work was also financially supported by 2021 Inamori Research Grants and the Information Center of Particle Technology.

\subsection*{Author contributions statement}
H.M. and K.S. contributed equally to this work.

%

%

\appendix

\section{Equation of motion~(\ref{eqmotionfr})}~\label{app.hessian}
In this section, we provide an explanation for the equation of motion~(\ref{eqmotionfr}), which includes the shear strain.
The particles, as well as the shear strain, are both displaced by $\ket{\vec{u}_\text{ex}} = \left[ \vec{u}_1,\cdots,\vec{u}_N,\sigma \epsilon \right]$~($dN+1$-dimensional vector) from their reference position $\ket{\vec{r}_\text{ex}} = \left[ \vec{r}_1,\cdots,\vec{r}_N,0 \right]$.
Previous studies~\cite{Tighe_2011,Baumgarten_2017} have established the overdamped equation of motion, whereas, in this work, we construct the underdamped equation of motion.

We begin with the equation of motion;
\begin{equation} \label{eqmotiontime}
m \frac{d^2 \ket{\vec{u}_\text{ex}}}{dt^2} = - \mathcal{M}_\text{ex} \ket{\vec{u}_\text{ex}} - \mathcal{C}_\text{ex} \frac{d\ket{\vec{u}_\text{ex}}}{dt} + \ket{\vec{F}},
\end{equation}
where $\mathcal{M}_\text{ex}$ and $\mathcal{C}_\text{ex}$ are the ``extended" Hessian and damping matrices, respectively.
We assign the mass $m$ of particles to the shear strain.
The external force is set to be $\ket{\vec{F}} = ({L^d}/{\sigma}) {\tau}_\text{sh} \ket{\gamma} = [\vec{0},\cdots,\vec{0}, (L^d/\sigma) \tau_\text{sh} ]$, where the shear stress $\tau_\text{sh}$ acts on the shear strain $\sigma \epsilon$.
Performing the Fourier transform on Eq.~(\ref{eqmotiontime}), we obtain Eq.~(\ref{eqmotionfr}).

Here we consider $xy$ shear strain, \textit{i.e.,} we align the $x$ axis with the displacement direction and the $y$ axis with the gradient direction.
When a strain $\epsilon$ is applied to the system, each particle $i$ undergoes an affine displacement of $\vec{u}^A_i = \epsilon y_i \vec{e}_x$, where $y_i$ represents the $y$ component of $\vec{r}_i$, and $\vec{e}_x$ is the unit vector in the $x$ direction.
Thus, we can define the ``extended" energy function as
\begin{equation}
\begin{aligned}
E_\text{ex} &= \sum_{i<j} \phi(r_{ij}), \\
r_{ij} &= |\vec{r}_i+\epsilon y_i \vec{e}_x -\vec{r}_j- \epsilon y_j \vec{e}_x|.
\end{aligned}
\end{equation}
Then, $\mathcal{M}_\text{ex}$ is defined as the second derivative of this function $E_\text{ex}(\vec{r}_1,\cdots, \vec{r}_N,\sigma\epsilon)$~(in the limit of $\sigma \epsilon \to 0$), which is expressed as follows;
\begin{equation} \label{eq:hessianex}
\mathcal{M}_\text{ex}
=
\left[
\begin{array}{cccc}
\mathcal{M}_{11} & \cdots & \mathcal{M}_{1N} & \displaystyle{ - \frac{\vec{\Sigma}_1}{\sigma} } \\
\vdots & \ddots & \vdots & \vdots \\
\mathcal{M}_{N1} & \cdots & \mathcal{M}_{NN} & \displaystyle{ - \frac{\vec{\Sigma}_N}{\sigma} } \\
\displaystyle{ - \frac{\vec{\Sigma}_1}{\sigma} } & \cdots & \displaystyle{ - \frac{\vec{\Sigma}_N}{\sigma} } & \displaystyle{ \frac{L^d}{\sigma^2}G_A }
\end{array}
\right],
\end{equation}
where $\mathcal{M}_{ij}={\partial^2 E_\text{ex}}/(\partial \vec{r}_i \partial \vec{r}_j )$~($d\times d$ matrix) are the same as components of the ``usual" Hessian matrix $\mathcal{M}$.
The vector $\vec{\Sigma}_i = -\partial^2 E_\text{ex} / (\partial \epsilon \partial \vec{r}_i)$ is the force acting on each particle $i$ which originates from an elementary affine deformation $\epsilon$, and $G_A$ represents the affine shear modulus~\cite{Mizuno3_2016}.

On the other hand, when we assume Stokes drag for the shear strain, we can define the ``extended" dissipation function as
\begin{equation}
R_\text{ex} = R(\vec{v}_1,\cdots, \vec{v}_N) + \frac{\zeta}{2} \left(\sigma \dot{\epsilon} \right)^2,
\end{equation}
where $R(\vec{v}_1,\cdots, \vec{v}_N)$ is the ``usual" dissipation function for the particles, $\vec{v}_i = d\vec{u}_i/dt$ is the velocity of particle $i$, and $\dot{\epsilon} = d\epsilon/dt$ is the rate of the shear strain.
Then, $\mathcal{C}_\text{ex}$ is defined as the second derivative of $R_\text{ex}(\vec{v}_1,\cdots, \vec{v}_N,\sigma\dot{\epsilon})$;
\begin{equation} \label{eq:dampingex}
\mathcal{C}_\text{ex}
=
\left[
\begin{array}{cccc}
\mathcal{C}_{11} & \cdots & \mathcal{C}_{1N} & 0 \\
\vdots & \ddots & \vdots & \vdots \\
\mathcal{C}_{N1} & \cdots & \mathcal{C}_{NN} & 0 \\
0 & \cdots & 0 & \zeta
\end{array}
\right],
\end{equation}
where $\mathcal{C}_{ij}={\partial^2 R}/(\partial \vec{v}_i \partial \vec{v}_j )$~($d\times d$ matrix) are components of the ``usual" damping matrix $\mathcal{C}$.
The present work assumes Stokes dissipation for particles, which is appropriate in both structural glasses and foams as discussed in Appendix~\ref{app.damping}.
In this case, the dissipation function is provided as
\begin{equation}
R_\text{ex} = \sum_{i} \frac{\zeta}{2} \vec{v}_i^2 + \frac{\zeta}{2} \left(\sigma \dot{\epsilon} \right)^2,
\end{equation}
and the damping matrix is $\mathcal{C}_\text{ex} = \zeta I_{dN+1}$.

By substituting $\mathcal{M}_\text{ex}$ in Eq.~(\ref{eq:hessianex}) and $\mathcal{C}_\text{ex}$ in Eq.~(\ref{eq:dampingex}) to Eq.~(\ref{eqmotiontime}), we get the equation of motion for $N$ particles as
\begin{equation} \label{eqshearapp1}
m \frac{d^2 \ket{\vec{u}}}{dt^2} = - \mathcal{M}\ket{\vec{u}} +  \epsilon \ket{\vec{\Sigma}} -\mathcal{C} \frac{d\ket{\vec{u}}}{dt},
\end{equation}
where $\ket{\vec{\Sigma}} = \left[ \vec{\Sigma}_1,\cdots,\vec{\Sigma}_N \right]$, and that for the shear strain as
\begin{equation} \label{eqshearapp2}
m \frac{d^2 (\sigma \epsilon)}{dt^2} = \frac{1}{\sigma}\bra{{\vec{\Sigma}}} \vec{u} \rangle - \frac{L^d}{\sigma^2}G_A (\sigma \epsilon) -\zeta \frac{d(\sigma \epsilon)}{dt} + \frac{L^d}{\sigma} \tau_\text{sh},
\end{equation}
or equivalently,
\begin{equation} \label{eqshearapp3}
\tau_\text{sh} = G_A \epsilon - \frac{1}{L^d} \bra{{\vec{\Sigma}}} \vec{u} \rangle + \frac{\sigma}{L^d} \left[ m \frac{d^2 (\sigma \epsilon)}{dt^2} + \zeta \frac{d(\sigma\epsilon)}{dt} \right].
\end{equation}
In addition, since $|\vec{u}_i| = \mathcal{O}(\sigma \epsilon)$ and $\bra{{\vec{\Sigma}}} \vec{u} \rangle = \mathcal{O}(L^d \epsilon)$, in the thermodynamic limit with $L \to \infty$~(and $N \to \infty$), Eq.~(\ref{eqshearapp3}) converges to
\begin{equation} \label{eqshearapp4}
\tau_\text{sh} = G_A \epsilon - \frac{1}{L^d} \bra{{\vec{\Sigma}}} \vec{u} \rangle,
\end{equation}
which means that the system is deformed in a quasi-static manner by the external shear stress $\tau_\text{sh}$. 
Thus, Eq.~(\ref{eqmotiontime}) or equivalently Eq.~(\ref{eqmotionfr}) in the main text can be expressed as Eq.~(\ref{eqshearapp1}) for particles and Eq.~(\ref{eqshearapp4}) for shear strain, separately.

When we perform the Fourier transform on Eqs.~(\ref{eqshearapp1}) and~(\ref{eqshearapp4}), we obtain
\begin{equation}
\begin{aligned}
-m\omega^2 \ket{\tilde{u}(\omega)} &= - \mathcal{M}\ket{\tilde{u}(\omega)} + \tilde{\epsilon}(\omega) \ket{\vec{\Sigma}}  - i \omega \mathcal{C} \ket{\tilde{u}(\omega)}, \\
\tilde{\tau}_\text{sh}(\omega) &= G_A \tilde{\epsilon}(\omega) - \frac{1}{L^d} \bra{\vec{\Sigma}} \tilde{u}(\omega) \rangle.
\end{aligned}
\end{equation}
By expanding the Hessian matrix $\mathcal{M}$ with its eigenvalues $m\omega_k^2$ and eigenvectors $\ket{\vec{e}_k}$ as $\mathcal{M} = \sum_{k=1}^{dN} m\omega_k^2 \ket{\vec{e}_k} \bra{\vec{e}_k}$ and supposing Stokes dissipation for the damping matrix as $\mathcal{C} = \zeta I_{2N}$, we obtain
\begin{equation} \label{eqoriginalmodulus}
\begin{aligned}
\ket{\tilde{u}(\omega)} &= \sum_{k=1}^{dN}\frac{ \ket{\vec{e}_k} \bra{ \vec{e}_k } \vec{\Sigma}\rangle }{m(\omega_k^2 - \omega^2) + i\omega\zeta } \tilde{\epsilon}(\omega),\\
\tilde{\tau}_\text{sh}(\omega) &= G_A \tilde{\epsilon}(\omega) - \frac{1}{L^d} \sum_{k=1}^{dN}\frac{ |\bra{ \vec{e}_k } \vec{\Sigma}\rangle|^2 }{m(\omega_k^2 - \omega^2) + i\omega\zeta } \tilde{\epsilon}(\omega),
\end{aligned}
\end{equation}
which are the same as formulations established in Ref.~\cite{Lemaitre_2006}.
The second equation in Eq.~(\ref{eqoriginalmodulus}) indicates the elastic modulus is composed of the affine $G_A$ and the non-affine $G_N(\omega)$ components as $\tilde{\tau}_\text{sh}(\omega) = [G_A - G_N(\omega)]\tilde{\epsilon}(\omega)$.

We mention the case of perfect crystals where the forces $\vec{\Sigma}_i$ for all particles~($i=1,\cdots,N$) become zero due to the symmetry of the lattice structures~\cite{Mizuno3_2016}.
The extended Hessian matrix becomes
\begin{equation} \label{eq:hessianexcrystal}
\mathcal{M}_\text{ex}
=
\left[
\begin{array}{cccc}
\mathcal{M}_{11} & \cdots & \mathcal{M}_{1N} & 0 \\
\vdots & \ddots & \vdots & \vdots \\
\mathcal{M}_{N1} & \cdots & \mathcal{M}_{NN} & 0 \\
0 & \cdots & 0 & \displaystyle{ \frac{L^d}{\sigma^2}G_A }
\end{array}
\right],
\end{equation}
which have $dN$ eigenvalues $m\omega_k^2$ and eigenvectors $\ket{\vec{e}_k}$ of the usual Hessain matrix $\mathcal{M}$ for $N$ particles, and one eigenvalue $L^d\sigma^{-2}G_A$ and eigenvector $\ket{\gamma}=[ \vec{0},\cdots,\vec{0}, 1]$ for the shear strain.
Thus, Eq.~(\ref{eqgastoriginal}) becomes
\begin{equation} \label{crystalmodulus}
{G_\ast (\omega)} =G_A +  \frac{\sigma^{2}}{ L^d } ( -m \omega^2 + i\omega \zeta ).
\end{equation}
In addition, Eqs.~(\ref{eqshearapp1}) and~(\ref{eqshearapp3}) become
\begin{equation} \label{crystalmodulus2}
\begin{aligned}
m \frac{d^2 \ket{\vec{u}}}{dt^2} &= - \mathcal{M}\ket{\vec{u}} -\mathcal{C} \frac{d\ket{\vec{u}}}{dt}, \\
\tau_\text{sh} &= G_A \epsilon + \frac{\sigma^2}{L^d} \left( m \frac{d^2 \epsilon}{dt^2} + \zeta \frac{d\epsilon}{dt} \right).
\end{aligned}
\end{equation}
The second equation in Eq.~(\ref{crystalmodulus2}) also gives Eq.~(\ref{crystalmodulus}).
In the crystalline case, particles and shear strain evolve independently of each other, and they do not interact.
The elastic modulus is characterized by the affine modulus $G_A$ only, and the non-affine modulus $G_N$ becomes zero.

\begin{figure*}[t]
\centering
\includegraphics[width=1.85\columnwidth]{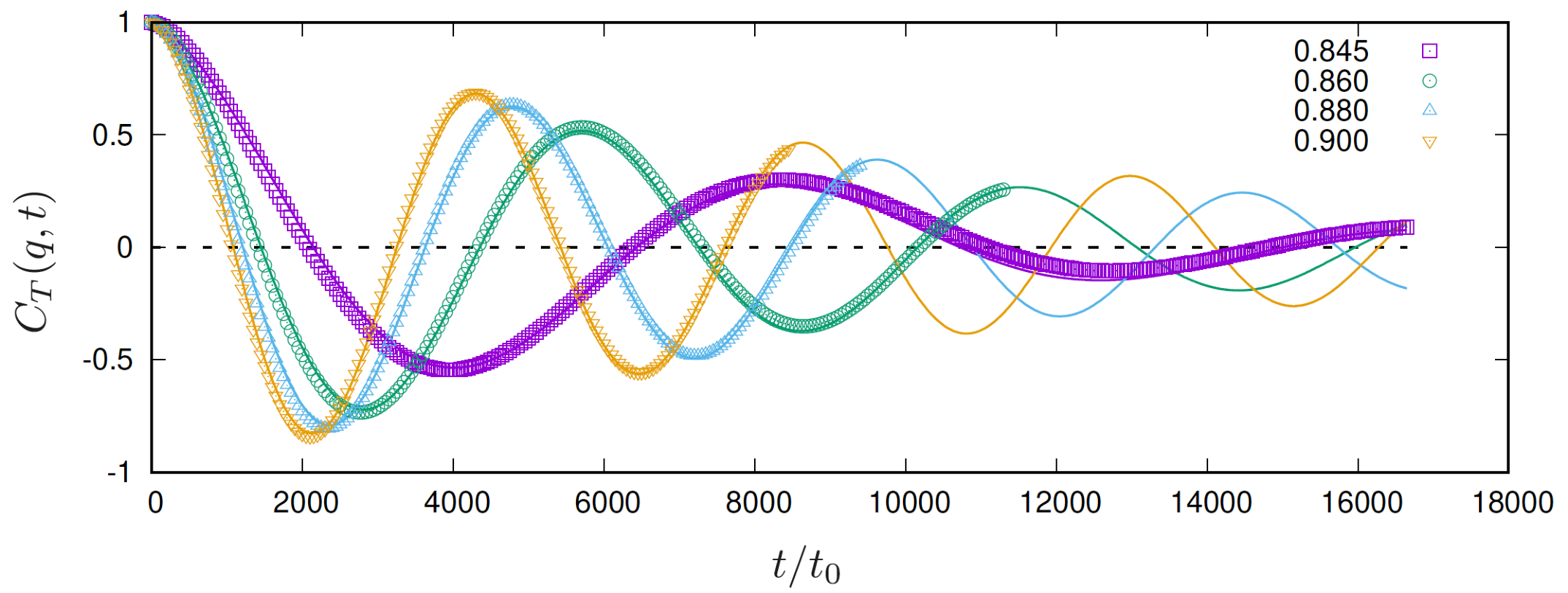}
\caption{ \label{fig:correlation}
The time correlation function of transverse velocity in the foams.
We plot $C_T(q,t)$ for various densities $\phi$ against $t/t_0$.
The symbols represent simulation data, and the solid lines show the damped harmonic oscillator model~(\ref{eq:dhom}) that is fitted to the simulation data.
The wave number is $q = 2\pi/L$~(where $L$ denotes the system length).
}
\end{figure*}

\begin{figure}[h]
\begin{center}
\includegraphics[width=\columnwidth]{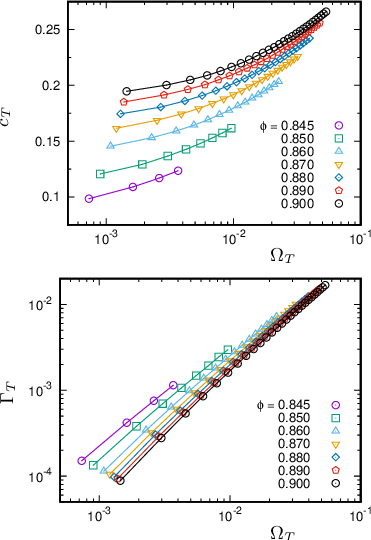}
\caption{ \label{fig:original}
Transverse shear waves in the foams.
We plot (a) $c_T$ and (b) $\Gamma_T$ as a function of $\Omega_T$.
The present data are scaled in Fig.~\ref{fig:originalscale} of the main text.
}
\end{center}
\end{figure}

\section{Damping matrix of foams}~\label{app.damping}
We demonstrate that the damping matrix of a two-dimensional~($d=2$) model of foams proposed by Durian~\cite{Durian_1995,Durian_1997} can be approximated to be Stokes dissipation as $\mathcal{C}_\text{ex} \approx \zeta I_{2N+1}$.
Since we suppose Stokes dissipation for the shear degree of freedom, we show that the ``usual" damping matrix $\mathcal{C}$ for $N$ particles can be approximated as $\mathcal{C} \approx \zeta I_{2N}$.

The damping matrix $\mathcal{C}$ is a $2N \times 2N$ matrix, and its $ij$ element $\mathcal{C}_{ij}$~($2 \times 2$ matrix) is given by
\begin{equation} \label{eq.damp}
\mathcal{C}_{ij}
=
\left\{ 
\begin{aligned}
& \mu z_i I_2 & (i=j), \\
& -\mu I_2 & (\text{particles $i$ and $j$ contact}),\\
& O & (\text{otherwise}),
\end{aligned} 
\right.
\end{equation}
where $z_i$ is the contact number of particle $i$, and $I_2$ is $2 \times 2$ unit matrix.
We expand $\mathcal{C}$ with eigenvectors $\ket{\vec{e}_k}$ of the Hessian matrix $\mathcal{M}$ as
\begin{equation}
\mathcal{C} = \sum_{k=1}^{2N} \sum_{l=1}^{2N} \ket{\vec{e}_k} \bra{\vec{e}_k} \mathcal{C}\ket{\vec{e}_l} \bra{\vec{e}_l}.
\end{equation}
Using Eq.~(\ref{eq.damp}) we obtain
\begin{equation}
\mathcal{C} = \sum_{k=1}^{2N} \sum_{l=1}^{2N} \ket{\vec{e}_k} \left\{ \sum_{i=1}^N \mu z_i \left( \vec{e}_{ki}\cdot \vec{e}_{li} \right) - \sum_{\left<ij\right>} \mu \left( \vec{e}_{ki}\cdot \vec{e}_{lj} \right) \right\}\bra{\vec{e}_l},
\end{equation}
where $\ket{\vec{e}_k} = \left[ \vec{e}_{k1},\vec{e}_{k2},\cdots,\vec{e}_{kN} \right]$, and $\sum_{\left<ij\right>}$ denotes summation over contacting pairs of particles $i$ and $j$.

We make the assumption that the fluctuations in the contact number are small, and therefore, the value of $z_i$ is approximately equal to the average contact number $z \approx 2d = 4$.
Additionally, we assume that the eigenvectors $\ket{\vec{e}_k}$ are random in nature and that $\vec{e}_{ki}\cdot \vec{e}_{lj}$ takes random values between $-1$ and $1$, such that $\sum_{\left<ij\right>} \vec{e}_{ki}\cdot \vec{e}_{lj} = 0$.
Based on these assumptions, we arrive at the equation;
\begin{equation} \label{eq.damp2}
\mathcal{C} \approx \sum_{k=1}^{2N} \mu z \ket{\vec{e}_k}  \bra{\vec{e}_k} = \mu z I_{2N},
\end{equation}
where we use the orthonormal condition $\sum_{i=1}^N \left( \vec{e}_{ki}\cdot \vec{e}_{li} \right) = \delta_{kl}$~($\delta_{kl}$ is the Kronecker delta).
Equation~(\ref{eq.damp2}) demonstrates Stokes dissipation, where $\mathcal{C} = \zeta I_{2N}$ with $\zeta = \mu z \approx 4\mu$.

\section{Overdamped dynamics for foams}~\label{app.over}
Here we consider the overdamped dynamics and determine the scaling laws of the complex shear modulus.
We start with the overdamped version of Eq.~(\ref{eqmotionfr});
\begin{equation}
\left( \mathcal{M}_\text{ex} + i\omega \mathcal{C}_\text{ex}\right) \ket{\tilde{u}_\text{ex}(\omega)} =  \frac{L^d}{\sigma} \tilde{\tau}_\text{sh}(\omega) \ket{\gamma}.
\end{equation}
$\mathcal{M}_\text{ex}$ is expanded with its eigenvalues $\lambda_{\text{ex} k}$~($\sim \omega_{\text{ex} k}^2$) and eigenvectors $\ket{\vec{e}_{\text{ex} k}}$ as $\mathcal{M}_\text{ex} = \sum_{k=1}^{dN+1} \lambda_{\text{ex} k} \ket{\vec{e}_{\text{ex} k}} \bra{\vec{e}_{\text{ex} k}}$, while $\mathcal{C}_\text{ex}$ is supposed to be Stokes dissipation as $\mathcal{C}_\text{ex} = \zeta I_{dN+1} = \sum_{k=1}^{dN+1} \zeta \ket{\vec{e}_{\text{ex} k}} \bra{\vec{e}_{\text{ex} k}}$.
We then get
\begin{equation}
\frac{1}{G_\ast (\omega)} = \frac{\sigma^{-1} \bra{\gamma} \tilde{u_\text{ex}}(\omega) \rangle }{\tilde{\tau}_\text{sh}(\omega)}
= \sum_{k=1}^{dN+1} \frac{L^d \sigma^{-2} |\bra{\vec{e}_{\text{ex} k}} \gamma \rangle|^2 }{\lambda_{\text{ex} k} + i\omega \zeta}.
\end{equation}
Assuming that $|\bra{\vec{e}_{\text{ex} k}} \gamma \rangle|$ does not depend on the eigenmodes $k$, we obtain
\begin{equation} \label{eqgast_over}
\frac{1}{G_\ast (\omega)} \sim \int_0^\infty d\lambda \frac{1 }{\lambda + i\omega \zeta} D(\lambda),
\end{equation}
where $D(\lambda)$ represents the vDOS like $g(\omega)$, but it describes distribution of the eigenvalues $\lambda_k$.
In Eq.~(\ref{eqgast_over}), we approxiamte $D_\text{ex}(\lambda)$ from $\mathcal{M}_\text{ex}$ by $D(\lambda)$ from $\mathcal{M}$.

Since $\lambda_k \sim \omega_k^2$, $D(\lambda) \sim \omega^{-1} g(\omega) \sim \lambda^{-1/2} g (\lambda^{1/2})$.
Using $g(\omega)$ in Eq.~(\ref{eq.dos}), we obtain
\begin{equation} \label{eq.dos_over}
D(\lambda)
\sim
\left\{ 
\begin{aligned}
& \lambda^{-1/2} & \left( \lambda_\ast < \lambda \right), \\
& \lambda^{-1/2} \left( \frac{\lambda}{\lambda_\ast} \right) & (\lambda_0 < \lambda < \lambda_\ast), \\
& \lambda^{-1/2} \left( \frac{\lambda}{\lambda_\ast} \right)^{(d+1)/2} & (\lambda < \lambda_0),
\end{aligned} 
\right.
\end{equation}
where $\lambda_\ast \sim \omega_\ast^2$ and $\lambda_0 \sim \omega_0^2$ are characteristic eigenvalues both of which follow the same scaling law as $\lambda_\ast \propto \lambda_0 \propto \Delta \phi \propto p$.

Substituting $D(\lambda)$ in Eq.~(\ref{eq.dos_over}) to $G_\ast(\omega)$ in Eq.~(\ref{eqgast_over}), we obtain the scaling laws of $G_\ast$ as
\begin{equation} \label{softvisco_over}
\frac{G_\ast}{\lambda_\ast^{1/2}}
\sim
\left\{
\begin{aligned}
& \left( \frac{\omega\zeta}{\lambda_\ast} \right)^{1/2} + i \left( \frac{\omega\zeta}{\lambda_\ast} \right)^{1/2} & \left( \frac{\omega\zeta}{\lambda_\ast} \gg 1 \right), \\
& 1 + i \left( \frac{\omega\zeta}{\lambda_\ast} \right)^{1/2} & \left( \frac{\lambda_0}{\lambda_\ast} \ll \frac{\omega\zeta}{\lambda_\ast} \ll 1 \right), \\
& 1 + i \left( \frac{\omega\zeta}{\lambda_\ast} \right) & \left(\frac{\omega\zeta}{\lambda_\ast} \ll \frac{\lambda_0}{\lambda_\ast} \right).
\end{aligned}
\right.
\end{equation}
When we substitute $\lambda_\ast = m \omega_\ast^2$ and $\lambda_0 = m \omega_0^2$, we can produce the same scaling laws of $G_\ast$ in Eq.~(\ref{softvisco}) of soft amorphous solids.

\section{Time correlation function of transverse velocity in foams}~\label{app.timecor}
We calculate the time correlation function $C_T(q,t)$ and use the damped harmonic oscillator model in Eq.~(\ref{eq:dhom}) to fit the time evolution of $C_T(q,t)$.
In Fig.~\ref{fig:correlation}, we include some examples of the fitting process, which shows that it works effectively.
This fitting process provides us with data on the frequency $\Omega_T(q)$ and the attenuation rate $\Gamma_T(q)$.

\section{Sound speed and attenuation rate in foams}~\label{app.sound}
Here we provide information about sound speed $c_T$ and attenuation rate $\Gamma_T$ as they vary with frequency $\Omega_T$.
The data in Fig.~\ref{fig:original} show the values for different densities $\phi$.
In the main text, we scale these data using the characteristic frequency $\omega_\ast \propto \Delta \phi^{1/2} \propto p^{1/2}$.
In Fig.~\ref{fig:originalscale}, we present the scaled values of $c_T/\omega_\ast^{1/2}$ and $\Gamma_T \zeta/(m\omega_\ast^2)$ as a function of $\Omega_T \zeta/(m\omega_\ast^2)$.
Regardless of the density $\phi$, these values collapse onto a single curve.

\bibliographystyle{apsrev4-2}
\bibliography{reference}
\end{document}